

\input jytex 


\typesize=12pt  \baselinestretch=1200
\sectionstyle{left}  \sectionnumstyle{blank}  \def\sec#1{{\bf\section{#1}}}
\def\eq#1{\eqno\eqnlabel{#1}$$}  \def\puteq#1{eq.~(\puteqn{#1})}
\def\ref#1{\markup{[\putref{#1}]}}
\def\to{\!\rightarrow\!}  \def\d{{\rm d}}
\def\bx{{\bf x}}  \def\Im{{\rm Im}}
\def\frac#1#2{{\textstyle {#1 \over #2}}}
\def\pmb#1{\setbox0=\hbox{#1}\kern-.01em\copy0\kern-\wd0  
  \kern.02em\copy0\kern-\wd0\kern-.01em\raise.017em\box0}


\pagenumstyle{blank}
\hbox to\hsize{\footnotesize\baselineskip=12pt
  \hfil\vtop{\hbox{\strut CALT-68-1797} \hbox{\strut DOE RESEARCH
  AND} \hbox{\strut DEVELOPMENT REPORT}}}
\vskip1.in
\centertext{\baselinestretch=1000\bigsize\bf
Complex Effective Potentials and Critical Bubbles%
\footnote{To appear in {\it Proceedings, Yale Workshop on Baryon Number
  Violation at the Electroweak Scale, March 1992\/} (World Scientific).
  Work supported in part by the U.S. Dept. of Energy under Contract
  No. DEAC-03-81ER40050.}}
\vskip .25in \footnotenumstyle{arabic}
\centertext{David E. Brahm%
\footnote[1]{Email: \tt brahm@theory3.caltech.edu}}
\vskip .2in \footnotenumstyle{symbols}
\centertext{\it California Institute of Technology, Pasadena, CA 91125}

\vskip .7in
\centerline{\bf Abstract}
\medskip {\narrower \baselinestretch=1100
The Higgs contribution to the effective potential appears to be complex.
How do we interpret this, and how should we modify the calculation to
calculate physical quantities such as the critical bubble free energy?
\par}

\vfill\lefttext{June, 1992}\medskip
\newpage \pagenum=0 \pagenumstyle{arabic}

\sec{1. A Toy Model with a Complex Effective Potential}
\hangindent=-2.3in \hangafter=3
\parshape=14 0in 6.5in 0in 6.5in 0in 3.7in 0in 3.7in 0in 3.7in 0in 3.7in
   0in 3.7in 0in 3.7in 0in 3.7in 0in 3.7in 0in 3.7in 0in 3.7in
   0in 3.7in 0in 6.5in
Suppose we wish to calculate the 1-loop finite-temperature effective
potential $V$ for a theory with a single scalar (I'll call it the Higgs),
whose tree potential $V_0$ (Fig.~1) is of the form
  $$V_0(\phi) = {\mu^2 \over 2\sigma^2} \phi^2 (\phi-\sigma)^2 -
    {\epsilon\phi^2 \over \sigma^2} \eq{V0}
The effective Higgs mass (at zero external momentum) is
  $$m^2(\phi,T) = V_0''(\phi) + {\mu^2\over 2\sigma^2} T^2 \eq{mh}
The last term of \puteq{mh} is the Higgs's ``self-plasma-mass'' (SPM); let
us choose our parameters to make the SPM small, so $m^2 < 0$ over roughly
$(1 \!-\! 1/\sqrt3)/2 < \phi/\sigma < (1 \!+\! 1/\sqrt3)/2$.

\hangindent=-2.3in \hangafter=17
The 1-loop contribution of the Higgs to $V$ can be calculated from the
vacuum-to-vacuum graph of Fig.~2a:
  $$V = V_0 + V_1 + {T^4\over2\pi^2} I(m/T) \approx V_0 + {T^2\over24}
    (V_0'') - {T\over12\pi} (V_0'')^{3/2} + \cdots \eq{vac}
or from the tadpole graph of Fig.~2b:
  $$V' = V_0' + V_1' + (V_0''') {T^2\over24} F(m/T) \approx V_0 +
    {T^2\over24} (V_0''') - {T\over8\pi} (V_0''') (V_0'')^{1/2} + \cdots
    \eq{tad}
where $V_1$ is the $T$-independent 1-loop result
  $$V_1 = {1\over4\pi^2} \int\!\d k \> k^2 \sqrt{k^2+m^2} = {m^4 \over
    64\pi^2} \left[ \ln\left( m^2 \over \Lambda^2 \right) - {3\over2}
    \right] \eq{V1}
and (writing $x=k/T$, $y=m/T$)
  $$I(y) \equiv \int_0^\infty\!\d x \> x^2 \ln\left( 1 - e^{-\sqrt{x^2+y^2}}
    \right), \qquad F(y) \equiv 6 I'(y)/(\pi^2 y) \eq{Idef}
I have expanded in small $m/T$.  Eqs.~\puteqn{vac} and \puteqn{tad}
give identical results.

\hangindent=-2.3in \hangafter=0
For $m^2 < 0$ the potential appears to be complex.  How are we to
interpret the imaginary part of the potential?  How should we modify $V$ to
get a real quantity to plot and use in calculations?  The naive answer,
which I'll call Method A, is simply to take the real part of $V$.  Several
much fancier methods can be found in the literature\ref{fancy}.

\sec{2. Relation to the Standard Model}
The toy model can, with minor modifications, represent the Standard Model
after integrating out gauge bosons and fermions.  Now $\{\mu,\,\sigma,\,
\epsilon\}$ depend on $T$ (and are simply related to the usual $\{
\lambda_T,\,E,\,D\}$\ref{AH}).  The contribution of the gauge bosons and
fermions to the Higgs plasma mass is still given correctly by \puteq{mh},
as can be verified by direct calculation of Feynman diagrams.  Goldstone
bosons double the SPM of \puteq{mh}, and themselves have a squared mass
$m_\chi^2 = V_0'/\phi + {\rm SPM}$ which becomes negative over $\frac12 <
\phi/\sigma < 1$ (for small SPM).  The tadpole calculation
\puteq{tad} must be used, using the 3-Higgs coupling of the original theory
($6\lambda\phi$ for a $\lambda\phi^4/4$ theory) in place of $(V_0''')$, to
avoid overcounting diagrams.  None of these modifications seem relevant to
the questions about imaginary parts.

\sec{3. Homogeneous and Inhomogeneous Fields}
At $T=0$, Weinberg and Wu showed that the imaginary part represents the
rate of decay of an unstable homogeneous field configuration to an
inhomogeneous state\ref{ww}.  Whether this is true at finite $T$ remains to
be shown.

For calculating percolation rates, however, we are more often interested in
the free energy $E_c$ of the critical bubble, an extremal configuration
stable against any fluctuation in $\phi(\bx)$ except overall growth or
shrinkage (the ``breathing mode''):
  $$E_c = \int\d^3\bx \left[ V(\phi(\bx)) + \frac12 (\nabla\phi)^2 +
    {AT\over m^3} \left( {\d m^2 \over \d\phi} \nabla\phi \right)^2 +
    {BT\over m^9} \left( {\d m^2 \over \d\phi} \nabla\phi \right)^4 +
    \cdots \right] \eq{ec}
Here $A,\,B\,\cdots$ come from derivative corrections to the
action\ref{chan,perry}.

\sec{4. \pmb{$\Im\{V\}$}\ Does Not Represent Bubble Growth/Shrinkage}
One might suppose that the contribution of $\Im\{V\}$ to $E_c$
represents the instability of the breathing mode.  We can disprove this
hypothesis by examining a thin-wall bubble [$\epsilon \ll \mu^2 \sigma^2/4$
in \puteq{V0}], for which\ref{fate}
  $$R = {2 S_1 \over\epsilon}, \qquad S_1 = \int\!\d\phi\sqrt{2V} =
    {2\mu \sigma^2 \over 9\sqrt3}, \qquad \delta = 1/\mu \eq{thin}
where $R$ is the bubble radius and $\delta$ is its thickness.
The contribution of $\Im\{V\}$ to $E_c$ is $\sim R^2$, since $V$ is only
complex in the bubble wall, and the wall profile is independent of
$\epsilon$ for $\epsilon\to0$.  The breathing mode imaginary contribution
to $E_c$, on the other hand, is independent of $R$, as can be
seen by calculating\ref{perry,esum,wass}
  $$E_c = E_0 + \sum_{n,l} (2l+1) \left[ {\omega_{n,l}\over2} + T \ln\left(
    1 - e^{-\omega_{n,l}/T} \right) \right] \eq{wsum}
where $E_0$ is the tree-level energy, and $\omega_{n,l}^2$ is the
eigenvalue of $[-\nabla^2 + V_0''(\phi(\bx))]$ whose eigenfunction has $n$
radial nodes and angular dependence $Y_l^m(\theta,\phi)$.
Eq.~\puteqn{wsum} is just the standard thermodynamic result for the free
energy of a system of harmonic oscillators.  The radial
part $\chi(r)/r$ of the eigenfunction satisfies
  $$\left[ {-\d^2\over\d r^2} + {l(l+1)\over r^2} + V_0''(\phi(r)) -
    \omega_{n,l}^2 \right] \chi_{n,l}(r) = 0 \eq{eigen}
and we see\ref{fate} that states bound to the wall approximately satisfy
  $$\omega_{n,l}^2 = \omega_{n,0}^2 + {l(l+1)\over R^2} \eq{nl}
The breathing mode eigenvalue $\omega_{0,0}^2$ can thus be found from the
translational mode eigenvalue $\omega_{0,1}^2 = 0$:
  $$\omega_{0,0}^2 = {-2 \over R^2}, \quad [E_c]_{0,0} \approx {i\over
    \sqrt2 R} + {i\pi T\over2} + T \ln\left( \sqrt2\over RT \right) \eq{w0}
One can argue\ref{wass} that \puteq{wsum} breaks down for unstable
fluctuations (inverted harmonic oscillators\ref{guth}), but even so it does
not appear that $\Im\{E_c\}$ grows as $R^2$.  Thus the contribution of
$\Im\{V\}$ to $E_c$ must be canceled by the derivative corrections of
\puteq{ec}.

Such cancellation is plausible, since odd powers of $m$ in the derivative
expansion give complex terms, and a similar cancellation is known to occur
to restore gauge invariance\ref{ginv,us}.  However, the divergences as $m
\to 0$ get increasingly worse, so this expansion seems inappropriate
for finding $\Im\{E_c\}$.

\sec{5. Removal of Long-Wavelength Modes}
The integral in \puteq{Idef} comes from a sum over Fourier modes of Higgs
field fluctuations, and the integrand is only complex for long wavelength
modes ($x<|y|$, or $k<|m|$).  At the hump ($\phi=\sigma/2$) for instance,
$m^2 = -\mu^2/2$, so only modes of wavelength $\lambda > 2\sqrt2\pi /
\mu$ contribute to $\Im\{V\}$.  This is several times the bubble wall
thickness\ref{GAco} $\delta=1/\mu$.

This suggests Method B\ref{us} for altering the calculation of $V$, namely
changing the lower limit of integration in \puteq{Idef}\ to $\Im\{y\}$.
Several schemes discussed in ref.~[\putref{fancy}] are in a similar spirit.
Note that Methods A and B are equivalent for the $T$-independent part
\puteq{V1}, but not for \puteq{Idef}, since the integrand of the latter in
the region $0 < x < \Im\{y\}$ is complex, not pure imaginary.

\sec{6. What's the Best Method?}
To decide on a ``best'' method of calculating physical quantities from a
complex $V$, we must decide what ``best'' means.  It could mean that when
we put our modified $V$ into \puteq{ec} and set $A=B=0$, we reproduce the
correct $E_c$.  Alternately, it could mean the method by which \puteq{tad}
gives the correct new degenerate minimum of $V$, as determined by
\puteq{vac} (Method A satisfies this criterion).

We usually bury our heads in the sand at this point, claiming the Higgs
sector contribution to $V$ is small in the Standard Model anyway.  As
experimental limits on the Higgs mass creep upward, however, it becomes
increasingly important to address these questions.


\sectionnumstyle{blank} \sec{References:}
\def\ap#1{{\it Ann.\ Phys.} {\bf #1}}
\def\pl#1{{\it Phys.\ Lett.} {\bf #1}}
\def\np#1{{\it Nucl.\ Phys.} {\bf #1}}
\def\pr#1{{\it Phys.\ Rev.} {\bf #1}}
\def\prl#1{{\it Phys.\ Rev.\ Lett.} {\bf #1}}

\def\zp#1{{\it Z.\ Phys.} {\bf #1}}
\baselinestretch=1000
\begin{putreferences}

\reference{fancy}{A. Okopinska, \pr{D36} (1987) 2415, \& refs.\ therein;\\
  I. Roditi, \pl{169B} (1986) 264;\\
  A. Ringwald \& C. Wetterich, \np{B334} (1990) 506.}
\reference{AH}{G.W. Anderson \& L.J. Hall, \pr{D45} (1992) 2685;\\
  M. Dine {\it et al.}, SLAC-PUB-5740 (Feb.\ 1992); SLAC-PUB-5741 (Mar.\
    1992).}
\reference{ww}{E.J. Weinberg \& A. Wu, \pr{D36} (1987) 2474.}
\reference{chan}{L.-H. Chan, \prl{54} (1985) 1222; {\bf 56} (1986) 404;\\
  C.M. Fraser, \zp{C28} (1985) 101;\\
  O. Cheyette, \prl{55} (1985) 2394.}
\reference{perry}{M. Li \& R.J. Perry, \pr{D37} (1988) 1670.}
\reference{fate}{S. Coleman, \pr{D15} (1977) 2929;\\
  C.G. Callan \& S. Coleman, \pr{D16} (1977) 1762.}
\reference{esum}{E.H. Wichmann \& N.M. Kroll, \pr{101} (1956) 843.}
\reference{wass}{D.A. Wasson \& S.E. Koonin, \pr{D43} (1991) 3400.}
\reference{guth}{A.H. Guth \& S.-Y. Pi, \pr{D32} (1985) 1899.}
\reference{ginv}{R. Kobes {\it et al.}, \prl{64} (1990) 2992; \np{B355}
    (1991) 1;\\
  I.J.R. Aitchison \& C.M. Fraser, \ap{156} (1984) 1.}
\reference{us}{D.E. Brahm \& S.D.H. Hsu, CALT-68-1705/HUTP-91-A063 (Dec.\
    1991);\\
  C.G. Boyd {\it et al.}, CALT-68-1795/HUTP-92-A027/EFI-92-22 (June 1992).}
\reference{GAco}{G. Anderson, private communication.}
\end{putreferences}
\bye